\documentclass{ws-p8-50x6-00}

\newcommand{\ba}[1]{\begin{eqnarray} \label{(#1)}}
\newcommand{\ea}{\end{eqnarray}}

\def\lsim{\mathrel{\vcenter{\hbox{$<$}\nointerlineskip\hbox{$\sim$}}}}
\def\gsim{\mathrel{\vcenter{\hbox{$>$}\nointerlineskip\hbox{$\sim$}}}}

\def \Rp{$R_p \hspace{-0.9em}/\;\:\hspace{-0.1em}$}
\def\egzk{E_{\rm GZK}}

\def\nue{\nu_e}

\def\nuebar{\bar{\nu}_e}

\def\dmsq{\delta m^2}

\begin{document}

\title{Absolute neutrino masses: physics beyond SM, double beta decay
and cosmic rays}

\author{Heinrich P\"as}

\address{Institut f\"ur Theoretische Physik und Astrophysik, Universit\"at 
W\"urzburg,\\ Am Hubland, 97074 W\"urzburg, Germany\\ 
E-mail: paes@physik.uni-wuerzburg.de}

\author{Thomas J. Weiler}

\address{Department of Physics and Astronomy, Vanderbilt University,\\
Nashville, TN 37235, USA \\
E-mail: tom.weiler@vanderbilt.edu}  


\maketitle

\abstracts{Absolute neutrino masses provide a key to physics beyond the 
standard model. We discuss the impact of absolute neutrinos masses on
physics beyond the standard model, the experimental possibilities to determine
absolute neutrinos masses, and the intriguing connection with the Z-burst 
model for extreme-energy cosmic rays.}

\section{Introduction}

Solar and atmospheric neutrino oscillations have established solid evidence 
for non-vanishing neutrino masses and determined   
mass squared differences $\dmsq$ and the mixing matrix $U$ with 
increasing accuracy. The puzzle of the absolute mass scale for
neutrinos, though, is still unsolved \cite{pw}.
In fact it is a true experimental challenge
to determine an absolute neutrino mass below 1~eV.
Three approaches 
have the potential to accomplish the task, namely  
larger versions of the tritium end-point distortion measurements,
limits from the evaluation of the large scale structure in the universe,
and 
next-generation neutrinoless double beta decay ($0\nu\beta\beta$) experiments.
In addition there is a fourth possibility: 
the extreme-energy cosmic-ray experiments 
in the context of the recently emphasized Z-burst model.
 
This article is organized as follows:
in section 2 the specific role of the neutrino among the elementary fermions
of the SM is reviewed,
as are the two most popular mechanisms for neutrino 
mass generation.  Also, 
the link of absolute neutrino masses to the physics 
underlying the standard model is discussed. Section 3 deals with direct 
determinations of the absolute neutrinos mass via tritium beta decay 
and cosmology.
In section 4 $0\nu\beta\beta$ is 
discussed, which may test very small values of neutrino masses, when  
information obtained from oscillation studies is input. Section 5 finally 
deals with the connection of the sub-eV neutrino mass scale and the ZeV 
energy scale of extreme energy cosmic rays in the Z-burst model. 

\section{Neutrino masses and physics beyond the standard model}
The specific role of the neutrino among the elementary fermions of the 
standard model (SM) is twofold: It is the only neutral particle, and its mass
is much smaller than the masses of the charged fermions. Thus it is 
self-evident that these properties may be related in a deeper theoretical 
framework underlying the standard model -- usually via Majorana 
mass-generating mechanisms.  
In the following we comment on the two most popular mechanisms to generate 
small neutrino masses, namely the see-saw mechanism and radiative neutrino 
mass generation.

\subsection{The see-saw mechanism}
\begin{figure}[t]
\begin{center}
\epsfxsize=20pc %
\epsfbox{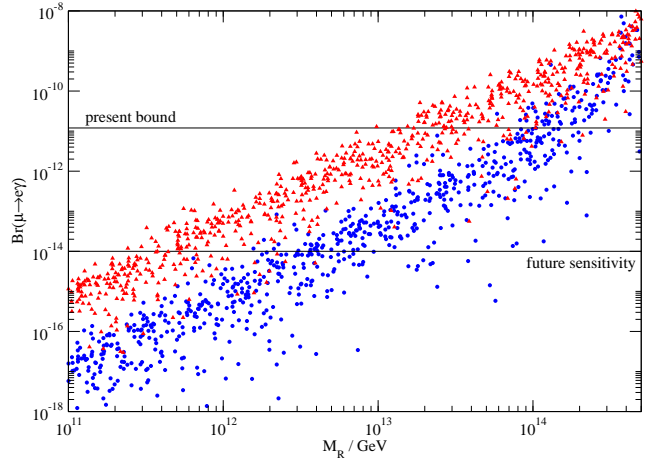}
\end{center} 
\caption{Branching ratio $\mu \rightarrow e \gamma$ as a function of
the right-handed Majorana mass $M_R$. Shown are large (lower curve)
and small (upper curve) neutrino mass scales, for a specific mSUGRA 
scenario. The scatter points correspond to estimated uncertainties in neutrino 
parameters after planned neutrino experiments have been performed.
  \label{depp}}
\end{figure}

The see-saw mechanism is based on the observation, that in order to generate
Dirac neutrino masses 
\be{}
m^D \overline{\nu^{}_L} \nu^{}_R
\label{dirac}
\ee
analogous to the mass terms of the charged leptons, the 
introduction of right-handed neutrinos is required. 
However a lepton-number violating
Majorana mass term 
for right-handed neutrinos
\be{}
\overline{\nu_R} M^R (\nu_R)^c
\label{maj}
\ee
is not prohibited by any gauge
symmetry of the standard model. Thus by buying 
a Dirac neutrino mass term $m^D$, one inevitably invites mixing with a  
Majorana mass $M^R \gg m^D$ 
which may live a priori at the mass scale of the underlying unified theory.
The diagonalization of the general mass matrix yields mass eigenvalues
\ba{}
m_{\nu} \simeq (m^D)^2/M^R \ll m^D 
\label{seesaw}
\\
M \simeq M^R, 
\ea
explaining the smallness of the light mass.
Though the fundamental scale $M^R$ is unaccessible for any kind of direct 
experimental testing,
it is obvious from eq. \ref{seesaw} 
that with information on the low-energy observables 
$m_{\nu}$ and $m^D$ the ``beyond the SM" mass scale of $M^R$ can be 
reconstructed. While it turns out to be unrealistic to determine $m^D$ in 
the standard model, this option exists indeed in supersymmetry.
In the  supersymmetric version of the see-saw mechanism,
lepton flavor violation (LFV) in the neutrino mass matrix 
(as required by neutrino oscillations) generates 
also LFV soft terms in the slepton mass matrix
proportional to the Dirac neutrino
Yukawa couplings \cite{hisano},
\be
\delta \tilde{m}_L^2 \propto Y_{\nu} Y_{\nu}^{\dagger}.
\ee
These LFV soft terms induce large branching ratios for SUSY mediated 
loop-decays such as 
$\mu \rightarrow e \gamma$,
\be
\Gamma(\mu \rightarrow e \gamma) \propto
\alpha^3 \frac{|(\delta \tilde{m}_L)^2_{ij}|^2}{m_S^8} \tan^2 \beta\,.
\ee
Here $m_S$ denotes the slepton mass scale in the loop.
Thus in the supersymmteric framework it is possible to probe the heavy mass 
scale $M_R$ by determining the (light) neutrino mass scale $m_\nu$
and the (Dirac) Yukawa couplings. This fact is illustrated in 
fig. \ref{depp}, where the branching ratio $\mu \rightarrow e \gamma$ 
is shown as a function of $M_R$
for a specific mSUGRA scenario, both for a large (lower curve)
and small (upper curve) neutrino mass scale \cite{dprr}. 

\subsection{Radiative neutrino masses}

\begin{figure}[t]
\begin{center}
\epsfxsize=20pc %
\epsfbox{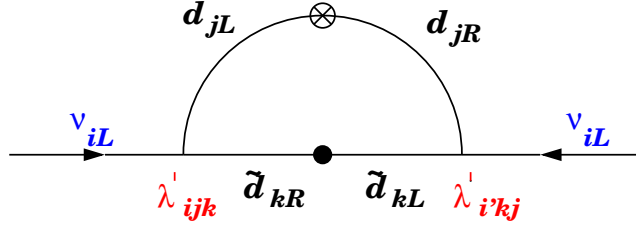}
\end{center} 
\caption{Radiative generation of neutrino Majorana masses in \Rp-violating
SUSY.
  \label{loop}}
\end{figure}

An alternative mechanism generates neutrino masses via loop graphs at the
SUSY scale, in contrast to the tree level generation of charged lepton masses
via the Higgs mechanism (see e.g. \cite{val}).
In supersymmetry lepton-number violating couplings
$\lambda$ and $\lambda'$ may arise if the discrete 
R-parity symmetry is broken (\Rp). These couplings may
induce neutrino masses via one loop self-energy graphs, see fig. \ref{loop}.
The entries in the neutrino mass matrix, given by
\be{}
m_{\nu_{ii'}} \simeq {{N_c \lambda'_{ijk} \lambda'_{i'kj}}
\over{16\pi^2}} m_{d_j} m_{d_k}
\left[\frac{f(m^2_{d_j}/m^2_{\tilde{d}_k})} {m_{\tilde{d}_k}} +
\frac{f(m^2_{d_k}/m^2_{\tilde{d}_j})} {m_{\tilde{d}_j}}\right],
\ee
are proportional to the products of \Rp-couplings and depending on the 
values of superpartner masses.
A determination of the absolute neutrinos mass scale would allow one to 
constrain all entries in the mass matrix, using the smallness of 
atmospheric and
solar $\dmsq$'s and unitarity of the neutrino mixing matrix $U$.
In fact, recent bounds on absolute neutrino masses improve
previous bounds on \Rp-couplings by up to 4 orders of magnitude \cite{bkp}.
Thus determining the neutrino mass probes physics at heavy 
mass scales beyond the SM also in the case of radiative generated neutrino 
masses.

\section{Absolute neutrino masses: direct determinations}

\subsection{Tritium beta decay}

In tritium decay, the larger the mass states comprising $\nuebar$,
the smaller is the Q-value of the decay.
The manifestation of neutrino mass is a reduction of phase space
for the produced electron at the high energy end of its spectrum.
An expansion of the decay rate formula about $m_{\nue}$ leads to
the end point sensitive factor 
\be{}
m^2_{\nu_e}\equiv \sum_j\,|U_{ej}|^2\,m^2_j\,,
\ee
where the sum is over mass states which can kinematically alter
the end-point spectrum.
If the neutrino masses are nearly degenerate,
then unitarity of $U$ leads immediately to a bound on
$\sqrt{m^2_{\nu_e}}=m_3$.
The design of a larger tritium decay experiment (KATRIN) to reduce 
the present 2.2~eV $m_{\nu_e}$ bound is under discussion;
direct mass limits as low as 0.4~eV, or even 0.2~eV, may be possible
in this type of experiment \cite{katrin}.

\subsection{CMB/LSS cosmological limits}

According to Big Bang cosmology, 
the masses of nonrelativistic neutrinos are related to the neutrino 
fraction of closure density by
$\sum_j m_j = 40\,\Omega_{\nu}\,h_{65}^2$~eV,
where $h_{65}$ is the present Hubble parameter in units of 65~km/s/Mpc.
As knowledge of large-scale structure (LSS) formation has evolved,
so have the theoretically preferred values for the hot dark matter (HDM)
component, $\Omega_\nu$.  In fact, the values have declined.
In the once popular HDM cosmology, one 
had $\Omega_\nu \sim 1$ and $m_\nu \sim 10$~eV 
for each of the mass-degenerate neutrinos.
In the cold-hot CHDM cosmology, the cold matter was dominant 
and one had $\Omega_\nu\sim 0.3$ and $m_\nu \sim 4$~eV
for each neutrino mass.
In the currently favored $\Lambda$DM cosmology,
there is scant room left for the neutrino component.
The power spectrum of early-Universe density perturbations
is processed by gravitational instabilities.
However, 
the free-streaming relativistic 
neutrinos suppress the growth of fluctuations
on scales below the horizon 
(approximately the Hubble size $c/H(z)$) 
until they become nonrelativistic at 
$z\sim m_j/3T_0 \sim 1000\,(m_j/{\rm eV})$.

A recent limit \cite{elg} 
derived from the 2dF Galaxy Redshift Survey power spectrum
constrains the fractional contribution of massive neutrinos to the total
mass density to be less than 0.13, translating into a bound on the
sum of neutrino mass eigenvalues,  $\sum_j m_j<1.8$~eV (for a total
matter mass density $0.1<\Omega_m<0.5$ and a scalar spectral index $n=1$).
Previous cosmological bounds come from the data of galaxy cluster abundances,
the Lyman $\alpha$ forest, data compilations of the cosmic microwave 
background (CMB), and peculiar 
velocities of large scale structure, and give upper bounds on the sum 
of neutrino masses in the range 3-6~eV.
A discussion of possible limits from future supernova neutrino detection
is given in \cite{BBM00}.

Some caution is warranted in the cosmological approach to neutrino mass,
in that the many cosmological parameters may conspire in 
various combinations to yield nearly identical CMB and LSS data.
An assortment of very detailed data may be needed to resolve 
the possible ``cosmic ambiguities''.

\section{Neutrinoless double beta decay}

$0\nu\beta\beta$
\cite{0vbb} corresponds to two single beta decays 
occurring simultaneously in 
one
nucleus and 
converts a nucleus (Z,A) into a nucleus (Z+2,A).
While even the standard model (SM) allowed process emitting two antineutrinos
\be
^{A}_{Z}X \rightarrow ^A_{Z+2}X + 2 e^- + 2 {\overline \nu_e}
\ee
is one of the rarest processes in nature with half lives in the region of
$10^{21-24}$ years, more interesting is the search for 
the neutrinoless mode,
\be        
^{A}_{Z}X \rightarrow ^A_{Z+2}X + 2 e^- 
\ee
which
violates lepton number by two units and thus implies physics beyond the 
SM. 

\begin{figure}[t]
\epsfxsize=10pc %
\begin{center} 
\epsfbox{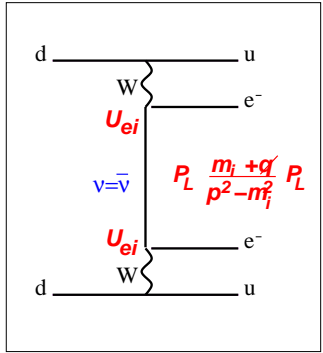}
\end{center} 
\caption{Diagram for neutrinoless double beta decay.
  \label{diag}}
\end{figure}

The $0\nu\beta\beta$ rate is a sensitive tool for the
measurement of the absolute mass-scale for Majorana neutrinos \cite{kps}.
The observable measured in the amplitude of $0\nu\beta\beta$ 
is the $ee$ element of the neutrino mass-matrix in the flavor basis
(see fig. \ref{diag}).
Expressed in terms of the mass eigenvalues and 
neutrino mixing-matrix elements, it is 
\be{}
m_{ee}= |\sum_i U_{ei}^2 m_i|\,.
\label{dbeqn}
\ee
A reach as low as $m_{ee}\sim 0.01$~eV seems possible 
with proposed double beta decay projects such as 
GENIUSI, MAJORANA, EXO, XMASS or MOON. 
This provides a substantial improvement over the current bound,
$m_{ee}< 0.6$~eV. A recent claim \cite{evi} 
by the Heidelberg-Moscow experiment
reports a best fit value of $m_{ee}=0.36$~eV,
but is subject to possible systematic uncertainties.
In the far future,
another order of magnitude in reach 
is available to the 
10 ton version of GENIUS, should it be funded and commissioned.

For masses in the interesting range $\gsim 0.01$~eV, 
the two light mass eigenstates are nearly degenerate and so the 
approximation $m_1 =m_2$ is justified.
Due to the restrictive CHOOZ bound, $|U_{e3}|^2 < 0.025$,
the contribution of the third mass eigenstate 
is nearly decoupled from $m_{ee}$ and so
$U^2_{e3}\,m_3$ may be neglected in the $0\nu\beta\beta$ formula.
We label by $\phi_{12}$ the relative phase between
$U^2_{e1}\,m_1$ and $U^2_{e2}\,m_2$.
Then, employing the above approximations,
we arrive at a very simplified expression for $m_{ee}$:
\be{}
m^2_{ee}=\left[1-\sin^2 (2\theta_{\rm sun})\,
       \sin^2 \left(\frac{\phi_{12}}{2}\right)\right]\,m^2_1\,.
\label{dbeqn2}
%
\ee
The two CP-conserving values of $\phi_{12}$ are 0 and $\pi$.
These same two values give maximal constructive and destructive
interference of the two dominant terms in eq.\ (\ref{dbeqn}),
which leads to upper and lower bounds for the observable
$m_{ee}$ in terms of a fixed value of $m_1$:
\be{}
\cos (2\theta_{\rm sun})\;m_1 \leq m_{ee} \leq m_1 \,,
\quad {\rm for\;\;fixed}\;\;m_1\,.
\label{dbbnds}
\ee
The upper bound becomes an equality, $m_{ee}=m_1$, if $\phi_{12}=0$.
The lower bound depends on Nature's value of the mixing angle
in the LMA solution.
A consequence of eq.\ (\ref{dbbnds}) is that for a given
measurement of $m_{ee}$, the corresponding inference of $m_1$ is 
uncertain over the range 
$[m_{ee},\,m_{ee}\,\cos (2\theta_{\rm sun})]$
due to the unknown phase difference $\phi_{12}$, with 
$\cos (2\theta_{\rm sun}) \gsim 0.1$ weakly bounded even 
for the LMA solution \cite{con}. This uncertainty disfavors $0\nu\beta\beta$
in comparison to direct measurements if a specific value of $m_1$
has to be 
determined, while $0\nu\beta\beta$ is more sensitive as long as 
bounds on $m_1$ are aimed at. 
Knowing the value of $\theta_{\rm sun}$ better will improve
the estimate of the inherent uncertainty in $m_1$.
For the LMA solar solution, the 
forthcoming Kamland experiment should reduce the error in the 
mixing angle $\sin^2 2 \theta_{\rm sun}$ to $\pm 0.1$ \cite{barger00}.

\section{Extreme energy cosmic rays in the Z-burst model}

It was expected that the EECR
primaries would be protons from outside the galaxy, produced in
Nature's most extreme environments such as the tori or radio hot spots 
of active galactic nuclei (AGN).
Indeed, cosmic ray data show a spectral flattening just below
$10^{19}$~eV
which can be interpreted as a new extragalactic component overtaking
the lower energy galactic component;
the energy of the break correlates well with the onset of a Larmor radius 
for protons too large to be contained by the Galactic magnetic field.
It was further expected that the extragalactic spectrum 
would reveal an end at the 
Greisen-Kuzmin-Zatsepin (GZK) cutoff energy of
$\egzk \sim 5\times 10^{19}$~eV. 
The origin of the GZK cutoff is the degradation of nucleon energy by the 
resonant scattering process $N+\gamma_{2.7K}\rightarrow \Delta^*
\rightarrow N+ \pi$ when the nucleon is above the resonant threshold $\egzk$.
The concomitant energy-loss factor is
$\sim (0.8)^{D/6 {\rm Mpc}}$ for a nucleon traversing a distance $D$. 
Since no AGN-like sources are known to exist within 100
Mpc of earth, the energy requirement for a proton arriving at earth with a
super-GZK energy is unrealistically high. 
Nevertheless, to date more than twenty events with energies 
at and above $10^{20}$~eV have been observed 
(for recent reviews see \cite{crrev}). \\

\begin{figure}[t]
\begin{center}
\epsfxsize=10pc %
\epsfbox{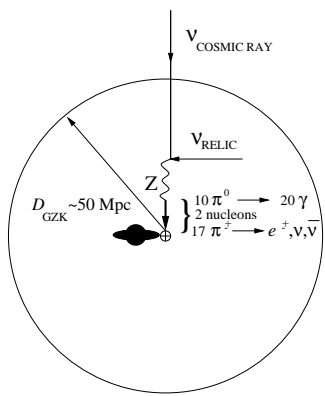}
\end{center} 
\caption{Schematic diagram showing the production of a Z-burst 
resulting from the 
resonant annihilation of a cosmic ray neutrino on a relic (anti)neutrino.
If the Z-burst occurs within the GZK zone ($\sim$ 50 to 100 Mpc) and is 
directed towards the earth, then photons and nucleons with energy above 
the GZK cutoff may arrive at earth and initiate super-GZK air-showers.
  \label{fig:nu2}}
\end{figure}

Several solutions have been proposed
for the origin of these EECRs,
ranging from unseen Zevatron accelerators (1~ZeV~$=10^{21}$~eV) 
and decaying supermassive particles and topological 
defects in the Galactic vicinity, 
to exotic primaries, exotic new interactions, and even exotic breakdown
of conventional physical laws.
A rather conservative and economical scenario involves cosmic ray 
neutrinos scattering resonantly on the cosmic neutrino background (CNB) 
predicted by Standard Cosmology, 
to produce Z-bosons \cite{Zburst}. 
These Z-bosons in turn decay to produce a highly boosted ``Z-burst'',
containing on average twenty photons and two nucleons above $\egzk$
(see Fig.\ 4).
The photons and nucleons from Z-bursts produced within 50 to 100 Mpc
of earth can reach earth with enough energy to initiate the 
air-showers observed at $\sim 10^{20}$~eV.

The energy of the neutrino annihilating at the peak of the Z-pole is
\be{}
E_{\nu_j}^R=\frac{M_Z^2}{2 m_j}=4\,(\frac{{\rm eV}}{m_j})\,{\rm ZeV}.
\ee

Even allowing for energy fluctuations about mean values, 
it is clear that in the Z-burst model the relevant
neutrino mass cannot exceed $\sim 1$~eV.
On the other hand, the neutrino mass cannot be too light
or the predicted primary energies will exceed the observed
event energies, and the primary neutrino flux will be pushed
to unattractively higher energies.
In this way,
one obtains a rough lower limit on the
neutrino mass of $\sim 0.1$~eV for the Z-burst model 
(with allowance made for an order of magnitude energy-loss 
for those secondaries traversing 50 to 100 Mpc). 
A detailed
comparison of the predicted proton spectrum with the observed EECR spectrum
in \cite{ringwald} yields a value of $m_{\nu}=2.34^{+1.29}_{-0.84}$~eV
for galactic halo and $m_{\nu}=0.26^{+0.20}_{-0.14}$~eV for extragalactic 
halo origin of the power-like part of the spectrum.

A necessary condition for the viability of this 
model is a sufficient flux of neutrinos at $\gsim 10^{21}$ eV.
Since this condition seems challenging, any increase of the Z-burst rate 
that ameliorates slightly the formidable flux requirement is helpful.
If the neutrinos are mass degenerate, then a further consequence is that
the Z-burst rate at $E_R$ is three times what it would be 
without degeneracy. 
If the neutrino is a Majorana particle,
a factor of two more is gained in the Z-burst rate relative 
to the Dirac neutrino case since the two active helicity states 
of the relativistic CNB
depolarize upon cooling to populate all spin states.

Moreover the viability of the Z-burst model is enhanced if the CNB neutrinos 
cluster in our matter-rich vicinity of the universe.
For smaller scales, the Pauli blocking of identical
neutrinos sets a limit on density enhancement.
As a crude
estimate of Pauli blocking, one may use the zero temperature Fermi gas as a
model of the gravitationally bound neutrinos. Requiring that the Fermi
momentum of the neutrinos does not exceed mass times the virial velocity 
$\sigma\sim\sqrt{MG/L}$ within the cluster of mass $M$ and size $L$, 
one gets the Tremaine-Gunn bound
\be{}
\frac{n_{\nu_j}}{54\,{\rm cm}^{-3}}\lsim 
10^3 \left(\frac{m_j}{{\rm eV}}\right)^3 
\left(\frac{\sigma}{200{\rm km/s}}\right)^3\,.
\ee
With a  virial velocity within our Galactic halo 
of a couple hundred km/sec
it appears that Pauli blocking allows significant clustering on the
scale of our Galactic halo only if $m_j \gsim 0.5$~eV.
Free-streaming (not considered here) also works against HDM clustering.

Thus, if the Z-burst model turns out to be the correct 
explanation of EECRs, then
it is probable that neutrinos possess one or more masses in the range 
$m_{\nu}\sim (0.1-1)$~eV. 
Mass-degenerate neutrino models are then likely. 
Some consequences are:

\begin{itemize}

\item
A value of $m_{ee}>0.01$~eV, and thus
a signal of $0\nu\beta\beta$ in next generation experiments,
assuming the neutrinos are Majorana particles.

\item
Neutrino mass sufficiently large to affect the 
CMB/LSS power spectrum. 

\end{itemize}

\section{Conclusions}

The absolute neutrino mass scale is a crucial parameter to 
learn about the theoretical structures underlying the SM\@.
Information about absolute neutrino masses can be obtained from direct
determinations via tritium beta decay or cosmology. More sensitive in giving 
limits but less valuable for determining the mass scale is neutrinoless 
double beta decay. The puzzle of EECRs above the GZK 
cutoff can be solved conservatively
with the Z-burst model, connecting the ZeV scale of EECRs 
to the sub-eV scale
of neutrino masses. If the Z-burst model turns out to be correct, 
neutrino masses in the region of 0.1-1~eV are predicted and  
degenerate scenarios are favored.
In this case positive signals for future tritium beta decay experiments, 
CMB/LSS studies, and $0\nu\beta\beta$ can be expected.

\section*{Acknowledgements}
HP would like to thank the organizers of NOON'01 for the kind invitation to 
this inspiring meeting. We also thank F. Deppisch for providing us with 
fig.~1.  
This work was supported by the
DOE grant no.\ DE-FG05-85ER40226 and the Bundesministerium 
f\"ur Bildung und Forschung (BMBF, Bonn, Germany) under the 
contract number 05HT1WWA2.

\end{document}